\documentclass[sigconf,natbib=true,nonacm]{acmart}

\usepackage{graphicx} 
\usepackage{xcolor}
\usepackage{xspace}
\usepackage{makecell}
\usepackage{balance}
\usepackage{array}

\newcommand{\pandasCorpus}{\textsc{PandasCorpus}\xspace}

\newcommand{\pandas}{\textsc{Pandas}\xspace}

\begin{document}

\title{PandasCorpus: A Resource of Real-World Pandas Workflows and Usage Patterns}

\author{Syrym Abdikhan}
\email{Syrym.Abdikhan@gisma-student.com}
\affiliation{%
  \institution{Gisma University of Applied Sciences}
  \country{Potsdam, Germany}
}

\author{Mazhar Hameed}
\email{mazhar.hameed@gisma.com}
\affiliation{%
  \institution{Gisma University of Applied Sciences}
  \country{Potsdam, Germany}
}

\begin{abstract}

\pandas has emerged as the de facto library for data processing and machine learning, widely used for tasks, such as data loading, transformation, and analysis. Despite its ubiquity, there has been limited systematic investigation into how \pandas is used in real-world projects and how typical workflows are composed in practice. To address this gap, we introduce \pandasCorpus, a dataset curated from GitHub repositories that captures real-world \pandas workflows at scale. In this work, a workflow refers to \pandas-based code contained in Jupyter notebooks, a prevalent medium for writing, executing, and sharing data analysis code.

The dataset comprises 139k notebooks from approximately 100k repositories and captures more than 4M \pandas API calls spanning 136 distinct operations. 
Beyond dataset construction, we characterize workflows using structural and \pandas-specific features and analyze notebook evolution between 2015 and 2025. Our study examines code executability, notebook size, and recurring sequences of \pandas operations, providing empirical insights into how \pandas is used in practice. The resulting corpus offers a reusable resource for studying data analysis workflows, \pandas usage patterns, and library-aware code composition. Both the dataset and the extraction pipeline are publicly available via GitHub\footnote{\label{note:github}\url{https://github.com/SyrymAbdikhan/PandasCorpus}} and Zenodo\footnote{\label{note:zenodo}\url{https://doi.org/10.5281/zenodo.17367371}}.

\end{abstract}


\maketitle

\section{\pandas as the De-Facto Library}
\label{sec:intro}

In recent years, data science and software engineering have increasingly relied on libraries that simplify and accelerate data manipulation~\cite{liao2020pandas,sdsu2022pythonss, macavaney2022python}. Among these, \pandas has become the de-facto standard for many data-driven workflows, offering a rich set of operations for data loading, transformation, and analysis~\cite{macdonald2018combining,hameed2024tasheeh,hameed2025repairing,rocklin2015dask}. While this flexibility empowers users, it also shifts substantial responsibility onto them, including how to compose operations effectively and efficiently. As a result, understanding how \pandas is used in real-world settings has become increasingly important.

This need is particularly pronounced in the context of computational notebooks, such as Jupyter~\cite{Jupyter}, which have emerged as the dominant medium for experimentation, teaching, and practical data science work. Jupyter notebooks are lightweight, easy to share, and support a wide ecosystem of libraries, making them widely adopted across academia and industry~\cite{DistilKaggle,tu2019parrot,sdsu2022pythonss, akkalyoncu2020lightweight}. Despite their popularity, our empirical understanding of how \pandas-based workflows are constructed and evolve within notebooks remains limited.

Prior research has analyzed Jupyter notebooks at scale, focusing on general aspects, such as code structure, reproducibility, and coding practices~\cite{DistilKaggle, Wang-icse, kgtorrent}. Similarly, other efforts have examined computational notebooks to understand machine learning practices, reproducibility challenges, and coding habits~\cite{yeye-Auto-pipeline,Wang-icse,yan-2020}.  Moreover, some studies have focused on specific libraries, such as NumPy~\cite{NumPy}, which is widely used for numerical computations~\cite{ziogas2021npbench,guelton2021numpyBenchmarks,guelton2025pythranTests}. While these efforts have provided valuable insights into notebook-based development, they do not focus on \pandas workflows specifically, nor do they examine how \pandas operations are composed and evolve in real-world notebook-based data analysis.

Analyzing \pandas-based workflows is critical for understanding how data manipulation tasks are structured in practice and for informing the design of data analysis tools and systems~\cite{mozzillo2023evaluation,elyashiv2024pd}. Although \pandas is widely adopted, its Pythonic execution model is not always optimized for performance, motivating extensive efforts to accelerate it through parallelization, distribution, or alternative execution backends, as seen in systems, such as Xorbits~\cite{lu2024Xorbits}, CuDF~\cite{cuDF}, Modin~\cite{modin1,modin2}, and Dias~\cite{dias}. Such efforts depend on representative, real-world \pandas code to guide system design and optimization. Beyond performance, an empirical understanding of \pandas operation usage and composition patterns is also essential for emerging applications, including automated data pipeline generation, benchmarking of data-processing systems, and the integration of large language models for code generation and transformation~\cite{vitagliano2023pollock,jindal2021magpie,modin1,dias}.

However, the absence of a dedicated resource capturing real-world \pandas workflows creates a significant gap. While platforms, such as GitHub host a vast number of Jupyter notebooks, these notebooks have not yet been systematically analyzed to provide \pandas-specific insights at scale. This limits our ability to answer fundamental questions about how \pandas is used in practice, which operations dominate real workloads, and how workflows evolve over time~\cite{modin2,mozzillo2023evaluation}.

To address this gap, we introduce \pandasCorpus, a resource of real-world \pandas workflows collected from GitHub repositories spanning the period from 2015 to 2025. The resource comprises 139k Jupyter notebooks from approximately 100k distinct repositories, organized into structured corpora and stored in Parquet format for efficient access. We characterize notebooks using general structural characteristics and \pandas-specific workflow characteristics, and conduct empirical analyses of notebook evolution and \pandas API usage, including temporal trends, code executability, and recurring operation sequences observed in practice.

\newpage

Our paper makes the following main contributions:
\begin{itemize}
    \item \pandasCorpus, a resource of real-world \pandas workflows collected from GitHub repositories, covering Jupyter notebooks created between 2015 and 2025.
    \item A reproducible pipeline for collecting, filtering, and organizing Jupyter notebooks containing \pandas-based workflows, enabling structured analysis at the notebook level.
    \item An empirical study of \pandas usage in practice, including dataset characterization, temporal trends in notebook evolution, and recurring operation and workflow patterns.
\end{itemize}

The remainder of this paper is organized as follows. Section~\ref{sec:pandasCorpus} introduces \pandasCorpus and describes the data collection and processing pipeline. Section~\ref{sec:findings} presents empirical insights into the dataset, focusing on notebook characteristics and \pandas usage patterns observed in practice. Section~\ref{sec:Implications} discusses potential use cases enabled by \pandasCorpus and concludes the paper.

\section{\LARGE \NoCaseChange{\pandasCorpus}}
\label{sec:pandasCorpus}

This section introduces \pandasCorpus and describes how the resource was constructed from GitHub repositories. We first outline the data collection and filtering process used to identify Jupyter notebooks containing \pandas-based workflows, and then describe how the resulting notebooks are organized and characterized.

\subsection{Process and Background}
\label{subsec:process&background}

\textbf{Data Source.} We collected data from public GitHub repositories using GitHub’s REST API~\cite{gapi}. The collection targeted Jupyter notebooks containing references to the \pandas API and was conducted in October 2025, covering repositories created between 2015 and 2025. To guide the data collection process, we selected 136 distinct \pandas API operations as search targets. These operations represent a comprehensive set of DataFrame-centric methods and functions that form the backbone of typical \pandas workflows. They include core manipulation operations (e.g., \texttt{drop}, \texttt{join}, \texttt{groupby}), data-cleaning steps (e.g., \texttt{fillna}, \texttt{replace}, \texttt{transform}), aggregation functions (e.g., \texttt{agg}, \texttt{sum}, \texttt{mean}), reshaping transformations, time-series operations, and common type-conversion utilities. Table~\ref{tab:crawl-stats} summarizes the overall statistics of \pandasCorpus. 
The collection complied with GitHub’s Terms of Service and excluded personally identifiable information.

\begin{table}
  \caption{\pandasCorpus Statistics}
  \small
  \label{tab:crawl-stats}
  \begin{tabular}{lc}
    \toprule
    \textbf{Name} & \textbf{Count} \\
    \midrule
    \pandas\ operations used for notebook crawling & 136 \\

    Total repositories crawled  & 137.1k \\
    Jupyter notebooks in crawled repositories~(25 GB) & 197.9k \\
   
    Repositories containing \pandas-based workflows & 99.9k \\
    Jupyter notebooks with \pandas-based workflows~(18 GB) & 139.0k  \\

    
    Repositories containing datasets~(e.g., CSV files) & 62.2k \\
    Code cells extracted & 3.8M \\
    Markdown cells extracted & 1.4M \\
    \bottomrule
  \end{tabular}
\end{table}

\textbf{Collection Pipeline.}
The collection pipeline consists of two stages: notebook discovery and content retrieval. We first generate query templates for 136 selected \pandas API operations combined with indicators of \pandas usage, such as import statements and file-reading operations. Queries are executed in both ascending and descending order to partially mitigate GitHub API limitations. The resulting notebook URLs are deduplicated, yielding approximately 198k candidate notebooks from 137k repositories. In the second stage, candidate notebooks are downloaded and filtered using the same relevance checks applied during discovery. This process results in 139k notebooks from approximately 100k repositories. 
Finally, the notebooks are organized into separate code-cell and markdown-cell corpora containing approximately 3.8M and 1.4M cells, respectively. While both corpora are retained, the analyses in this study focus on code cells. The dataset is stored in compressed Parquet format for efficient storage and access.

\textbf{Code Parsing and Characterization.}
We analyze individual code cells to characterize \pandas usage within notebooks. Before parsing, code is sanitized by removing IPython magic and shell commands and resolving minor syntax inconsistencies. The cleaned code is then parsed into an Abstract Syntax Tree (AST)\footnote{\url{https://docs.python.org/3/library/ast.html}}, enabling precise identification of function calls and syntactic constructs. We traverse the AST to extract notebook- and workflow-level characteristics. Compared to substring-based matching, AST-based analysis reduces false positives by distinguishing executable code from comments and string literals, although occasional overlaps with non-\pandas objects may still occur.

\subsection{Workflow Characterization Dimensions}
\label{subsec:feature_matrix}

We organize the extracted information into two groups of workflow characteristics: general notebook-level characteristics and \pandas-specific workflow characteristics. The first group captures structural properties of notebooks, including code size, organization, and complexity, while the second focuses on \pandas API usage and workflow composition patterns.

\textbf{General Notebook-Level Characteristics.}
The general notebook-level characteristics summarized in Table~\ref{tab:Characteristics-all} are commonly used in prior studies on computational notebooks~\cite{kgtorrent,DistilKaggle,dyer2013boa,Wang-icse}. We retain these established characteristics to provide a structural baseline and enable comparison with existing notebook corpora, rather than as \pandas-specific descriptors. The results show substantial variation in notebook size and structure. The median notebook contains 139 lines of code and exhibits a nested block depth of two, suggesting moderate structural complexity. The distributions are right-skewed, with a subset of notebooks containing thousands of lines or statements, reflecting the heterogeneity of coding practices across Jupyter notebooks. Building on this general characterization, \pandasCorpus additionally captures \pandas-specific usage and composition characteristics that are not represented by these conventional notebook-level measures.

\begin{table}[!ht]
\caption{Statistics of General Notebook-Level and \pandas-Specific Workflow Characteristics}
\label{tab:Characteristics-all}
\small
\begin{tabular}{rlccc}
  \toprule
  \textbf{\#} & \textbf{Characteristic} & \textbf{\#Notebooks}  & \textbf{Med} & \textbf{Max} \\
  \midrule\midrule
  & \multicolumn{4}{l}{Notebook-Level Characteristics} \\
  \midrule\midrule
1  & Lines of Code                      & 139,032     & 139    & 6,764 \\
2  & Number of Blank Lines of Code      & 127,215     & 17     & 1,229 \\
3  & Number of Statements               & 139,032     & 86     & 3,909 \\
4  & Number of Parameters               & 77,225      & 1      & 1,254 \\
5  & Number of User-Defined Functions   & 78,272      & 1      & 325   \\
6  & Nested Block Depth                 & 107,522     & 2      & 18    \\
7  & Number of Operands                 & 139,030     & 85     & 9,615 \\
8  & Number of Operators                & 139,027     & 51     & 6,176 \\
9  & Number of Unique Operands          & 139,030     & 13     & 3,421 \\
10 & Number of Unique Operators         & 139,027     & 4      & 30    \\
11 & Number of Identifiers              & 139,032     & 134    & 5,207 \\
12 & Number of Imports                  & 138,932     & 8      & 608   \\
  \midrule\midrule
  & \multicolumn{4}{l}{\pandas-Specific Characteristics} \\
  \midrule\midrule
1  & Number of Pandas Operations        & 139,032     & 18     & 2,918 \\
2  & Longest Chained Operation          & 139,032     & 2      & 95    \\
3  & Number of Chained Operations       & 139,032     & 16     & 2,112 \\
4  & Total Length of Chained Operations & 139,032     & 18     & 2,918 \\
5  & Longest Chained Indexing           & 136,960     & 1      & 13    \\
6  & Number of Chained Indexing         & 136,960     & 24     & 4,169 \\
7  & Total Length of Chained Indexing   & 136,960     & 25     & 5,129 \\
8  & Number of Visualizations           & 59,542      & 0      & 234   \\
9  & Number of Training Calls           & 57,322      & 0      & 91    \\
10 & Number of Shell Commands & 39,738      & 0      & 261   \\
  \bottomrule
\end{tabular}
\end{table}

\textbf{\pandas-Specific Workflow Characteristics.}
While the general characteristics describe the structure of computational notebooks, they do not capture how data-manipulation operations are used and composed within them. To provide a \pandas-oriented representation of real-world workflows, \pandasCorpus therefore augments these conventional notebook characteristics with a set of \pandas-specific characteristics capturing API usage frequency, chained operations, chained indexing, visualization activity, model-training calls, and the use of IPython magic and shell commands. Chained operations describe compositions of multiple \pandas calls within a single expression (e.g., \texttt{df.groupby().agg().reset\_index()}), while chained indexing captures successive indexing operations (e.g., \texttt{df[mask][col]}). Table~\ref{tab:Characteristics-all} summarizes the resulting characteristics and their statistics.
Notebook-level statistics for these characteristics provide an overview of how \pandas is used in practice. Each notebook contains at least one \pandas operation, with a median of 18 operations and an interquartile range between 9 and 35. The usage distribution is highly skewed, with a small number of notebooks containing nearly 3,000 \pandas operations. However, the median longest chained operation is 2, indicating that multi-step operation chaining is common across notebooks.

These characteristics provide a descriptive framework for analyzing notebook structure and \pandas usage. In the following section, we examine notebook executability, temporal trends, \pandas operation usage, and workflow composition patterns.

\section{Analysis and Findings}
\label{sec:findings}

Building on the characterization of \pandasCorpus presented in Section~\ref{sec:pandasCorpus}, we now examine how \pandas workflows manifest in practice. This section investigates notebook executability over time, common error types, the evolution of \pandas API usage, and recurring operation sequences observed across the dataset.

\subsection{Notebook Executability Over Time}
\label{subsec:notebookExecutability}

\begin{figure}[!h]
  \centering
  \includegraphics[width=0.97\linewidth]{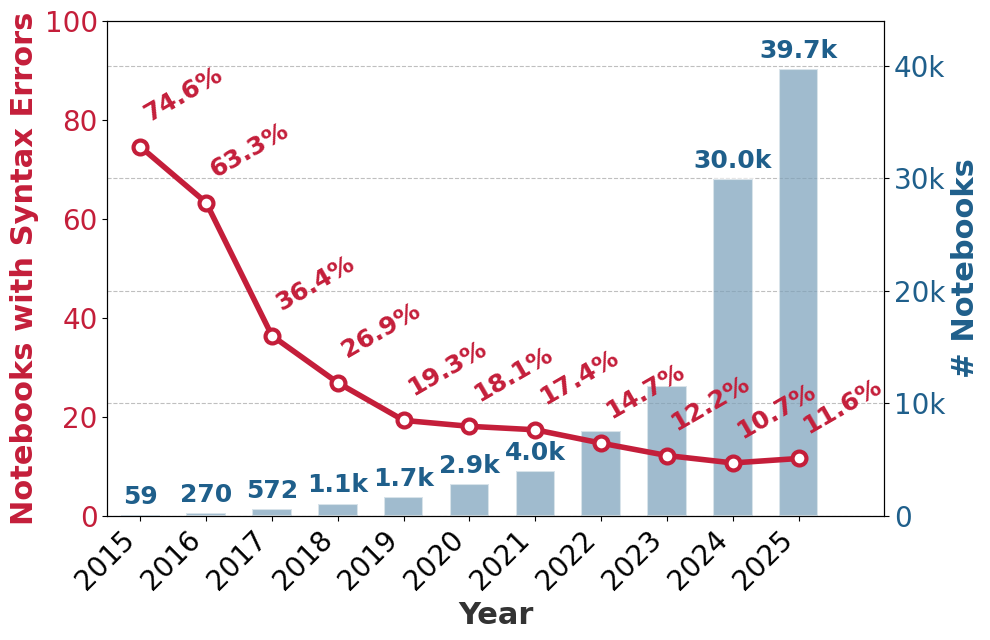}
  \caption{Notebooks Containing Syntax Errors Over Time}
  \label{fig:error_otime}
\end{figure}

Figure~\ref{fig:error_otime} shows the percentage of notebooks containing at least one syntax-level error, as detected through AST parsing, over the period 2015–2025. Errors were identified when the Python \texttt{ast} module raised a \texttt{SyntaxError} during parsing. Since the AST parser halts at the first syntax violation, the reported values reflect notebooks containing at least one syntactic inconsistency.

The results indicate a substantial decrease in syntax-level errors over time. In 2015, approximately 74.6\% of notebooks contained syntax errors. This percentage decreases steadily over time, reaching 10.7\% in 2024, followed by a slight increase to 11.6\% in 2025. Overall, the long-term trajectory indicates substantial improvement in notebook syntactic validity. At the same time, the number of notebooks grows dramatically, particularly after 2022, increasing from fewer than 5k notebooks per year before 2021 to 30k in 2024 and nearly 40k in 2025. Despite this rapid growth, the relative error rate continues to decline, suggesting that increased volume does not correspond to increased syntactic instability.

A pronounced drop is observed between 2016 and 2017, where the error rate decreases from 63.3\% to 36.4\%. While the dataset does not allow us to attribute this shift to a single cause, it may reflect broader ecosystem changes, such as maturation of notebook tooling, evolution of Python versions, or improved development practices. Similarly, the continued decline after 2022 coincides with a period of rapid growth in notebook creation, though causal interpretation requires further investigation.

To better understand the nature of syntax failures, Table~\ref{tab:error_cat_2} summarizes the distribution of error categories extracted from parsing errors. The majority of errors are structural syntax issues, including unmatched or unclosed brackets (42.07\%), unterminated strings (21.05\%), and generic syntax errors (16.51\%). Indentation-related errors account for an additional 10.4\%. Since errors are detected statically at parse time, these categories reflect syntactic inconsistencies rather than runtime or semantic failures. The dominance of bracket- and string-related errors suggests that many notebooks contain incomplete or partially written code fragments, which is consistent with the iterative and exploratory nature of real-world notebook-based development.

\begin{table}[!h]
\caption{Syntax Error Categories}
\label{tab:error_cat_2}
\begin{tabular}{lc}
  \toprule
  \textbf{Error Category} & \textbf{Percentage} \\
  \midrule
  Unmatched or unclosed brackets & 42.07\% \\
  Unterminated string literal
        & 21.05\% \\
  Generic syntax error     & 16.51\% \\
  Indentation-related error       & 10.4\%  \\
  Missing comma in expression      & 3.87\%  \\
 Invalid decimal literal         & 3.32\%  \\
 Invalid or unexpected character       & 1.99\%  \\
  Other                      & 0.77\%  \\
  \bottomrule
\end{tabular}

\end{table}

\begin{table*}[]
\centering

\caption{Frequency of 136 \pandas operations across \pandasCorpus, ranked by usage. Counts represent the number of observed operation calls, and percentages indicate their share of all recorded \pandas calls.}
\label{tab:pandas_freq_fullset_new}

\setlength{\tabcolsep}{5pt}
\begin{tabular}{
r l r r
@{\hspace{8pt}}|@{\hspace{8pt}}
r l r r
@{\hspace{8pt}}|@{\hspace{8pt}}
r l r r
}
\toprule
\textbf{\#} & \textbf{Operation} & \textbf{Count} & \textbf{\%} &
\textbf{\#} & \textbf{Operation} & \textbf{Count} & \textbf{\%} &
\textbf{\#} & \textbf{Operation} & \textbf{Count} & \textbf{\%} \\
\midrule

1  & head             & 300877 & 7.50 & 47 & sort\_index       & 19860 & 0.49 & 93  & cumprod             & 3960 & 0.10 \\
2  & mean             & 223797 & 5.58 & 48 & get\_dummies      & 18046 & 0.45 & 94  & take                & 3898 & 0.10 \\
3  & sum              & 202306 & 5.04 & 49 & to\_numeric       & 17354 & 0.43 & 95  & aggregate           & 3752 & 0.09 \\
4  & groupby          & 198799 & 4.95 & 50 & resample          & 16508 & 0.41 & 96  & rename\_axis        & 3710 & 0.09 \\
5  & drop             & 156530 & 3.90 & 51 & duplicated       & 16004 & 0.40 & 97  & droplevel           & 3594 & 0.09 \\
6  & apply            & 144117 & 3.59 & 52 & assign           & 15432 & 0.38 & 98  & add\_prefix         & 3537 & 0.09 \\
7  & astype           & 135149 & 3.37 & 53 & unstack          & 14286 & 0.36 & 99  & factorize           & 3305 & 0.08 \\
8  & reset\_index     & 132761 & 3.31 & 54 & pivot\_table     & 14092 & 0.35 & 100 & tz\_convert         & 3207 & 0.08 \\
9  & join             & 115077 & 2.87 & 55 & update           & 13372 & 0.33 & 101 & between\_time       & 3158 & 0.08 \\
10 & replace          & 102480 & 2.55 & 56 & cumsum           & 13175 & 0.33 & 102 & sem                 & 3083 & 0.08 \\
11 & value\_counts    & 100374 & 2.50 & 57 & squeeze          & 12551 & 0.31 & 103 & product             & 2958 & 0.07 \\
12 & sort\_values     & 99509  & 2.48 & 58 & mode             & 12540 & 0.31 & 104 & prod                & 2838 & 0.07 \\
13 & copy             & 88413  & 2.20 & 59 & diff             & 12519 & 0.31 & 105 & kurtosis            & 2760 & 0.07 \\
14 & unique           & 87775  & 2.19 & 60 & stack            & 12295 & 0.31 & 106 & add\_suffix         & 2656 & 0.07 \\
15 & concat           & 87511  & 2.18 & 61 & transpose        & 12145 & 0.30 & 107 & nsmallest           & 2616 & 0.07 \\
16 & fillna           & 85858  & 2.14 & 62 & filter           & 11734 & 0.29 & 108 & corrwith            & 2590 & 0.06 \\
17 & merge            & 84685  & 2.11 & 63 & eval             & 11062 & 0.28 & 109 & combine\_first      & 2575 & 0.06 \\
18 & rename           & 72700  & 1.81 & 64 & insert           & 10980 & 0.27 & 110 & kurt                & 2502 & 0.06 \\
19 & dropna           & 72697  & 1.81 & 65 & idxmax           & 10972 & 0.27 & 111 & to\_timestamp       & 2480 & 0.06 \\
20 & max              & 70321  & 1.75 & 66 & notna            & 10936 & 0.27 & 112 & expanding           & 2414 & 0.06 \\
21 & isnull           & 59656  & 1.49 & 67 & reindex          & 9695  & 0.24 & 113 & bfill               & 2395 & 0.06 \\
22 & get              & 52841  & 1.32 & 68 & notnull          & 9380  & 0.23 & 114 & set\_axis           & 2223 & 0.06 \\
23 & isin             & 50553  & 1.26 & 69 & pct\_change      & 9286  & 0.23 & 115 & swaplevel           & 1962 & 0.05 \\
24 & transform        & 49427  & 1.23 & 70 & var              & 8907  & 0.22 & 116 & merge\_asof         & 1855 & 0.05 \\
25 & set\_index       & 48745  & 1.21 & 71 & pop              & 8709  & 0.22 & 117 & convert\_dtypes     & 1825 & 0.05 \\
26 & min              & 48635  & 1.21 & 72 & cut              & 8654  & 0.22 & 118 & cummax              & 1773 & 0.04 \\
27 & describe         & 48367  & 1.21 & 73 & crosstab         & 8591  & 0.21 & 119 & combine             & 1674 & 0.04 \\
28 & map              & 47434  & 1.18 & 74 & nlargest         & 8131  & 0.20 & 120 & align               & 1461 & 0.04 \\
29 & std              & 42621  & 1.06 & 75 & skew             & 7991  & 0.20 & 121 & truncate            & 1412 & 0.04 \\
30 & isna             & 40075  & 1.00 & 76 & clip             & 7850  & 0.20 & 122 & compare             & 1310 & 0.03 \\
31 & where            & 35487  & 0.88 & 77 & melt             & 7690  & 0.19 & 123 & to\_xarray          & 1308 & 0.03 \\
32 & count            & 35448  & 0.88 & 78 & pivot            & 7489  & 0.19 & 124 & at\_time            & 1285 & 0.03 \\
33 & agg              & 35098  & 0.87 & 79 & to\_period       & 6146  & 0.15 & 125 & first\_valid\_index & 1136 & 0.03 \\
34 & round            & 32721  & 0.82 & 80 & rank             & 5797  & 0.14 & 126 & last\_valid\_index  & 822  & 0.02 \\
35 & drop\_duplicates & 29131  & 0.73 & 81 & explode          & 5756  & 0.14 & 127 & reorder\_levels     & 762  & 0.02 \\
36 & shift            & 28018  & 0.70 & 82 & ffill            & 5683  & 0.14 & 128 & wide\_to\_long      & 733  & 0.02 \\
37 & median           & 26052  & 0.65 & 83 & idxmin           & 5642  & 0.14 & 129 & merge\_ordered      & 610  & 0.02 \\
38 & sample           & 23959  & 0.60 & 84 & ewm              & 5455  & 0.14 & 130 & reindex\_like       & 510  & 0.01 \\
39 & nunique          & 23825  & 0.59 & 85 & pipe             & 5049  & 0.13 & 131 & cummin              & 490  & 0.01 \\
40 & to\_numpy        & 23540  & 0.59 & 86 & interpolate      & 4609  & 0.11 & 132 & infer\_objects      & 457  & 0.01 \\
41 & quantile         & 23406  & 0.58 & 87 & cov              & 4346  & 0.11 & 133 & asof                & 424  & 0.01 \\
42 & rolling          & 22915  & 0.57 & 88 & asfreq           & 4256  & 0.11 & 134 & T                   & 51   & $<$0.01 \\
43 & query            & 22896  & 0.57 & 89 & tz\_localize     & 4165  & 0.10 & 135 & from\_dummies       & 50   & $<$0.01 \\
44 & tail             & 22221  & 0.55 & 90 & qcut             & 4149  & 0.10 & 136 & lreshape            & 19   & $<$0.01 \\
45 & abs              & 21237  & 0.53 & 91 & mask             & 4118  & 0.10 &     &                     &      &       \\
46 & corr             & 20878  & 0.52 & 92 & xs               & 4073  & 0.10 &     &                     &      &       \\
\bottomrule
\end{tabular}
\end{table*}

\subsection{\pandas Operations Coverage}
\label{subsec:pandasOperations}

The \pandas API provides a broad collection of operations for data manipulation, aggregation, reshaping, and transformation. Identifying which of these operations are most frequently used helps characterize practical usage patterns and uncover the core building blocks of real-world workflows. To examine how \pandas is used in practice, we analyze the frequency of 136 DataFrame-centric operations across all notebooks in the corpus. Table~\ref{tab:pandas_freq_fullset_new} presents all 136 \pandas DataFrame operations ranked in descending order by usage frequency.

\begin{figure}
  \centering
  \includegraphics[width=0.98\linewidth]{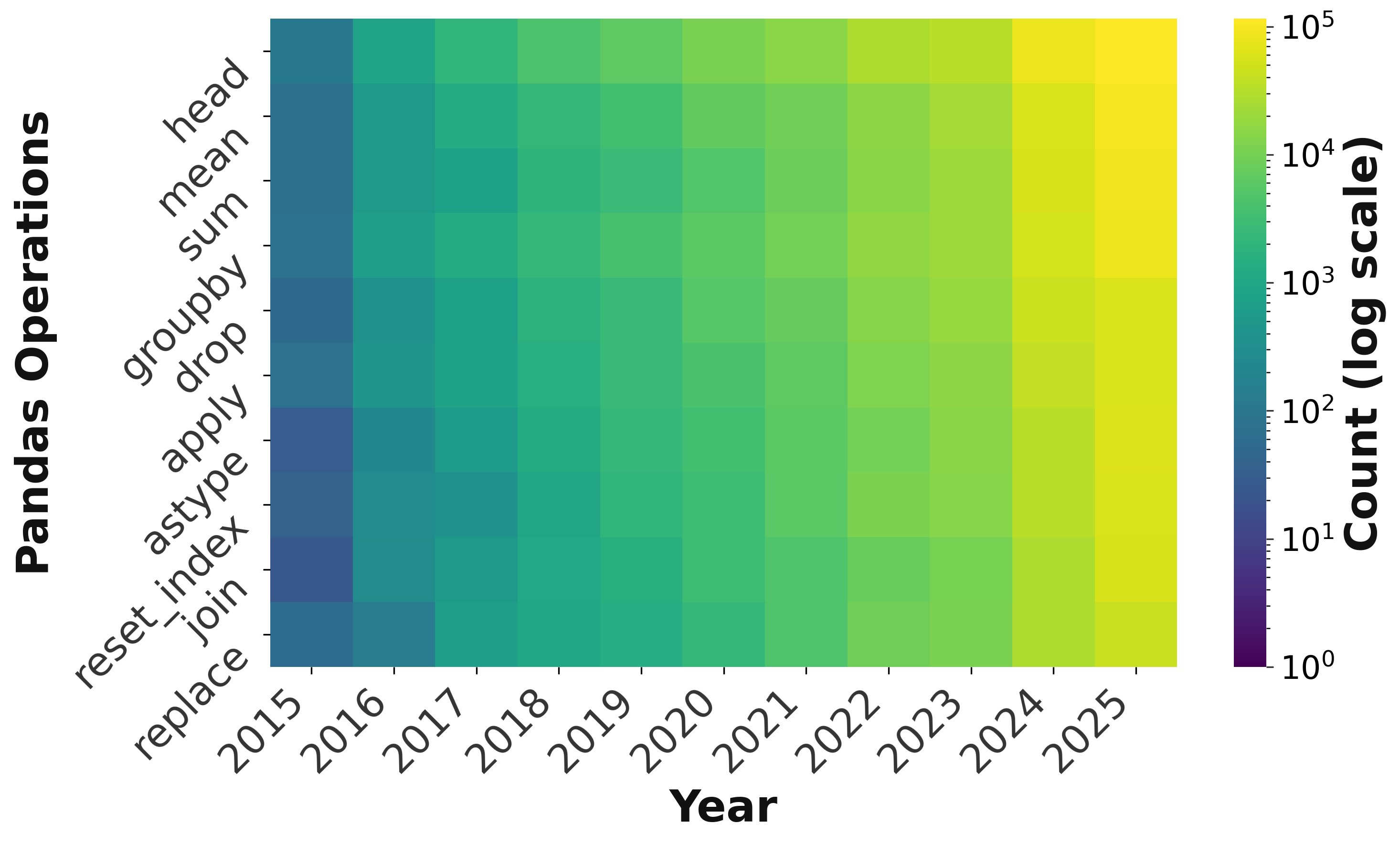}
  \caption{Temporal evolution of the top 10 \pandas operations (ranked by overall usage).}
  \label{fig:top_pandas_otime}
  
\end{figure}

Figure~\ref{fig:top_pandas_otime} illustrates the usage trends of the most frequently invoked operations from 2015 to 2025 (log-scaled counts). The distribution is highly skewed. A small subset of core operations dominates usage throughout the entire period. In total, the top 15 operations account for more than half of all recorded \pandas calls, while the least-used operations contribute negligibly to overall usage.

Core manipulation and aggregation operations, such as \texttt{head}, \texttt{mean}, \texttt{sum}, \texttt{groupby}, and \texttt{drop} consistently rank among the most frequent across all years. Their steady growth over time reflects both the increasing volume of notebooks and the central role of these operations in typical workflows. Less frequently used operations, including specialized reshaping or time-related transformations, appear sporadically and at much lower magnitudes. This long-tail pattern suggests that real-world \pandas workflows are built around a compact set of recurring transformation primitives rather than the full breadth of the API. Figure~\ref{fig:top_pandas_otime} illustrates the distribution of selected operations, highlighting the dominance of a small subset of functions, while Table~\ref{tab:pandas_freq_fullset_new} provides the complete ranked list of all 136 \pandas DataFrame operations.

\subsection{Composition Patterns of Pandas Operations}
\label{subsec:pandasOperationSEQ}

\begin{table}
\caption{Most Frequent Pandas Operation Sequences}

\label{tab:sequence}
\begin{tabular}{rlc}
  \toprule
  \textbf{\#} & \textbf{Operation Sequence} & \textbf{\%} \\
  
  \midrule
  \multicolumn{3}{c}{Integration \& Construction} \\
  \midrule
  1 & \texttt{join → read\_csv → join} & 2.7\% \\
  2 & \texttt{DataFrame → DataFrame → concat} & 2.17\% \\
  3 & \texttt{read\_csv → read\_csv → concat} & 2.04\% \\
  4 & \texttt{head → merge → head} & 2.03\% \\
  5 & \texttt{read\_csv → read\_csv → merge} & 1.92\% \\
  6 & \texttt{read\_csv → join → read\_csv} & 1.84\% \\
  7 & \texttt{DataFrame → DataFrame → merge} & 1.58\% \\
  8 & \texttt{read\_csv → join → read\_csv → join} & 1.57\% \\
  9 & \texttt{DataFrame → to\_csv → read\_csv} & 1.43\% \\
  
  \midrule
  \multicolumn{3}{c}{Exploration \& Validation} \\
  \midrule
  1  & \texttt{head → isnull → sum} & 4.25\% \\
  2  & \texttt{head → sort\_values → head} & 3.29\% \\
  3  & \texttt{head → drop → head} & 3.22\% \\
  4  & \texttt{read\_csv → head → describe} & 2.6\% \\
  5  & \texttt{sort\_values → head → sort\_values} & 2.26\% \\
  6  & \texttt{read\_csv → head → isnull} & 2.29\% \\
  7  & \texttt{read\_csv → head → tail} & 2.16\% \\
  8  & \texttt{read\_csv → head → info} & 2.12\% \\
  9  & \texttt{read\_csv → head → drop} & 2.12\% \\
  10 & \texttt{sum → duplicated → sum} & 2.09\% \\
  11 & \texttt{read\_csv → head → isnull → sum} & 1.86\% \\
  12 & \texttt{sort\_values → head → sort\_values → head} & 1.71\% \\
  13 & \texttt{read\_csv → head → value\_counts} & 1.4\% \\
  14 & \texttt{head → unique → unique} & 1.32\% \\
  
  \midrule
  \multicolumn{3}{c}{Cleaning \& Transformation} \\
  \midrule
  1 & \texttt{fillna → mean → fillna} & 1.95\% \\
  2 & \texttt{isnull → sum → dropna} & 1.91\% \\
  3 & \texttt{duplicated → sum → drop\_duplicates} & 1.08\% \\
  4 & \texttt{read\_csv → to\_datetime → set\_index} & 1.12\% \\

  \midrule 
  \multicolumn{3}{c}{Aggregation \& Statistics} \\
  \midrule
  1 & \texttt{groupby → agg → reset\_index} & 4.86\% \\
  2 & \texttt{groupby → mean → groupby} & 3.12\% \\
  3 & \texttt{rolling → mean → rolling} & 3.11\% \\
  4 & \texttt{mean → rolling → mean} & 2.58\% \\
  5 & \texttt{mean → groupby → mean} & 2.46\% \\
  6 & \texttt{sum → sum → mean} & 1.45\% \\
  7 & \texttt{groupby → sum → reset\_index} & 1.42\% \\
  8 & \texttt{groupby → mean → reset\_index} & 1.35\% \\
  9 & \texttt{groupby → apply → reset\_index} & 1.31\% \\
  \bottomrule
\end{tabular}
\end{table}

While frequency analysis reveals which operations dominate individually, it does not capture how operations are composed in practice. To better understand workflow structure, we analyze recurring sequences of \pandas API calls observed across notebooks. These sequences reflect higher-level composition patterns that characterize real-world \pandas workflows. To identify such patterns, we extract sequences of executable \pandas API calls by analyzing control-flow paths within notebooks. Only operations reachable through actual execution paths are considered. 

To focus on meaningful patterns, we apply three filtering criteria:
(1)~sequences must have a minimum length of two operations;
(2)~sequences must appear in at least 1\% of notebooks ($\approx$1,400 notebooks);
(3)~trivial repetitions of the same operation (e.g., repeated \texttt{mean} or \texttt{apply}) are excluded. The resulting sequences, summarized in Table~\ref{tab:sequence}, are grouped into four categories reflecting common workflow stages: integration and construction, exploration and validation, cleaning and transformation, and aggregation and statistics. Overall, the identified sequences reveal that \pandas workflows exhibit recurring structural patterns rather than arbitrary compositions. This consistency suggests that real-world \pandas usage is organized around a limited set of stable workflow motifs.

\textbf{Integration \& Construction.} Sequences in this category are characterized by repeated occurrences of \texttt{read\_csv}, \texttt{join}, \texttt{merge}, and \texttt{concat}, indicating that data loading and integration form a central stage of many workflows. Patterns, such as \texttt{read\_csv → concat} and \texttt{read\_csv → merge} suggest that analysts frequently combine multiple external data sources early in the workflow. The interleaving of inspection operations (e.g., \texttt{head → merge → head}) indicates that integration steps are often validated immediately after execution. In addition, sequences, such as \texttt{DataFrame → to\_csv → read\_csv} reveal intermediate materialization and re-loading of data, possibly for debugging or restructuring purposes.

\textbf{Exploration \& Validation.} Sequences in this category highlights the importance of quick visual inspection in notebook-based workflows. Typical patterns, such as \texttt{read\_csv → head → describe} and \texttt{read\_csv → head → info} indicate that data exploration commonly begins immediately after loading a dataset. Validation-related sequences, including \texttt{head → isnull → sum} and \texttt{sum → duplicated → sum}, suggest systematic checks for missing values and duplicate records. These patterns reflect standard data quality verification steps embedded directly into exploratory workflows. The frequent alternation between sorting and inspection (e.g., \texttt{sort\_values → head}) further illustrates the iterative nature of exploration, where users repeatedly refine views of the data.

\textbf{Cleaning \& Transformation.} 
Sequences in this category reveal structured preprocessing behavior preceding analysis. Patterns, such as \texttt{isnull → sum → dropna} and \texttt{duplicated → sum → drop\_duplicates} indicate that cleaning operations are typically preceded by explicit validation steps, suggesting that users first quantify data quality issues before modifying the dataset. The sequence \texttt{fillna → mean → fillna} reflects staged imputation strategies, where missing values are first partially addressed, summary statistics are computed, and then used to perform statistically informed imputation on remaining fields. Additionally, the pattern \texttt{read\_csv → to\_datetime → set\_index} reflects a canonical preprocessing pipeline for time-series data, where temporal columns are parsed and promoted to an index for subsequent analysis.

\textbf{Aggregation \& Statistics.}  Sequences in this category are strongly centered around \texttt{groupby}, which appears in the majority of the most frequent patterns. The canonical structure \texttt{groupby → agg → reset\_index} (4.86\%) highlights a standard workflow in which grouped summaries are materialized back into flat tabular form.  Several additional patterns, such as \texttt{groupby → mean → groupby} and \texttt{mean → groupby → mean}, indicate iterative re-grouping and multi-stage aggregation, suggesting that users often refine summaries across multiple grouping dimensions.

Taken together, these categories indicate that \pandas workflows often exhibit a recurring progression. While not strictly linear in every notebook, this layered structure reflects a compositional regularity in real-world \pandas usage and provides a foundation for future workflow modeling, optimization, and automated assistance.

\section{Implications and Conclusion}
\label{sec:Implications}

In this paper, we introduced \pandasCorpus, a resource of real-world \pandas workflows collected from GitHub repositories spanning 2015 to 2025. The corpus comprises 139k Jupyter notebooks from approximately 100k repositories and provides a structured basis for studying how \pandas is used in practice. Through empirical analyses, we examined notebook executability trends over time, the frequency of commonly used \pandas operations, and recurring operation compositions observed in real-world workflows.

\pandasCorpus enables a range of research directions centered around real-world data manipulation workflows. First, the corpus facilitates code and workflow retrieval. Unlike traditional code datasets that focus on isolated functions, \pandasCorpus preserves workflow structure, enabling retrieval models to consider sequences of operations and their contextual dependencies. This supports code search, workflow recommendation, and next-step prediction in interactive data analysis environments. Second, the dataset enables mining and modeling of recurring workflow patterns. The extracted operation sequences reveal compositional regularities that can inform automated pipeline construction, workflow abstraction, and intent modeling. Finally, \pandasCorpus supports program analysis and optimization research. The corpus provides a large-scale collection of real-world \pandas workflows suitable for studying refactoring strategies, detecting anti-patterns, improving performance, and evaluating library-aware tooling.

\balance
\bibliographystyle{ACM-Reference-Format}
\bibliography{bibliography}

@misc{gapi,
  title = {GitHub REST API Documentation},
  year = {2025},
  howpublished = {\url{https://docs.github.com/en/rest?apiVersion=2022-11-28}},
  note = {Accessed date: 2025-07-28}
}

@misc{ast,
  title = {Abstract Syntax Trees — Python 3.13.7 documentation},
  year = {2025},
  howpublished = {\url{https://docs.python.org/3/library/ast.html}},
}

@article{DistilKaggle,
	author = {Ghahfarokhi, Mojtaba Mostafavi and Asgari, Arash and Abolnejadian, Mohammad and Heydarnoori, Abbas},
	journal = {Proceedings of the 21st International Conference on Mining Software Repositories (MSR)},
	pages = {647--651},
	title = {{DistilKaggle: A Distilled Dataset of Kaggle Jupyter Notebooks}},
	year = {2024},
}

@article{yan-2020,
	author = {Yan, Cong and He, Yeye},
	journal = {Proceedings of the 2020 ACM SIGMOD International Conference on Management of Data},
	title = {{Auto-Suggest: Learning-to-Recommend Data Preparation Steps Using Data Science Notebooks}},
	year = {2020},
}

@article{kgtorrent,
  title={Kgtorrent: A dataset of python jupyter notebooks from kaggle},
  author={Quaranta, Luigi and Calefato, Fabio and Lanubile, Filippo},
  journal={In 18th International Conference on Mining Software Repositories (MSR)},
  pages={550--554},
  year={2021}
}

@article{modin2,
  title={Flexible rule-based decomposition and metadata independence in modin: a parallel dataframe system},
  author={Petersohn, Devin and Tang, Dixin and Durrani, Rehan and Melik-Adamyan, Areg and Gonzalez, Joseph E and Joseph, Anthony D and Parameswaran, Aditya G},
  journal={Proceedings of the VLDB Endowment},
  volume={15},
  number={3},
  year={2021}
}

@article{modin1,
title = {Towards scalable dataframe systems},
author = {Petersohn, Devin and Macke, Stephen and Xin, Doris and Ma, William and Lee, Doris and Mo, Xiangxi and Gonzalez, Joseph E. and Hellerstein, Joseph M. and Joseph, Anthony D. and Parameswaran, Aditya},
 journal={Proceedings of the VLDB Endowment},
volume = {13},
number = {12},
year = {2020}}

@article{dias,
  title={Dias: Dynamic Rewriting of Pandas Code},
  author={Baziotis, Stefanos and Kang, Daniel and Mendis, Charith},
  journal={Proceedings of the ACM on Management of Data},
  volume={2},
  number={1},
  pages={1--27},
  year={2024},
  publisher={ACM New York, NY, USA}
}

@inproceedings{dyer2013boa,
  title={Boa: A language and infrastructure for analyzing ultra-large-scale software repositories},
  author={Dyer, Robert and Nguyen, Hoan Anh and Rajan, Hridesh and Nguyen, Tien N},
  booktitle={35th International Conference on Software Engineering (ICSE)},
  pages={422--431},
  year={2013},
  organization={IEEE}
}

@misc{liao2020pandas,
  author       = {Li-Ting Liao},
  title        = {Solving Real-World Business Questions with Python Pandas},
  year         = {2020},
  howpublished = {\url{https://medium.com/li-ting-liao-tiffany/solving-real-world-business-questions-with-pandas-70ef8ef02675}}
}

@misc{sdsu2022pythonss,
  author       = {{San Diego State University}},
  title        = {Python for Social Scientists, Linguistics/BDA 572},
  year         = {2022},
  howpublished = {\url{https://gawron.sdsu.edu/python_for_ss}}
}

@inproceedings{jindal2021magpie,
  title={Magpie: Python at Speed and Scale using Cloud Backends},
  author={Jindal, Alekh and Emani, K Venkatesh and Daum, Maureen and Poppe, Olga and Haynes, Brandon and Pavlenko, Anna and Gupta, Ayushi and Ramachandra, Karthik and Curino, Carlo and Mueller, Andreas and others},
  booktitle={Conference on Innovative Data Systems Research (CIDR)},
  year={2021}
}

@inproceedings{rocklin2015dask,
  title={Dask: Parallel computation with blocked algorithms and task scheduling.},
  author={Rocklin, Matthew and others},
  booktitle={SciPy},
  pages={126--132},
  year={2015}
}

@article{vitagliano2023pollock,
  title={Pollock: A data loading benchmark},
  author={Vitagliano, Gerardo and Hameed, Mazhar and Jiang, Lan and Reisener, Lucas and Wu, Eugene and Naumann, Felix},
  journal={Proceedings of the VLDB Endowment},
  volume={16},
  number={8},
  pages={1870--1882},
  year={2023}
}

@article{hameed2025repairing,
  title={Repairing Raw Data Files with TASHEEH},
  author={Hameed, Mazhar and Vitagliano, Gerardo and Panse, Fabian and Naumann, Felix},
  journal={ACM SIGMOD Record},
  volume={54},
  number={1},
  pages={90--99},
  year={2025}
}

@article{hameed2024tasheeh,
  title={Tasheeh: Repairing row-structure in raw csv files},
  author={Hameed, Mazhar and Vitagliano, Gerardo and Panse, Fabian and Naumann, Felix},
  journal={International Conference on Extending Database Technology (EDBT)},
  pages={426--439},
  year={2024}
}

@inproceedings{ziogas2021npbench,
  title={NPBench: A benchmarking suite for high-performance NumPy},
  author={Ziogas, Alexandros Nikolaos and Ben-Nun, Tal and Schneider, Timo and Hoefler, Torsten},
  booktitle={Proceedings of the 35th ACM International Conference on Supercomputing},
  pages={63--74},
  year={2021}
}

@misc{guelton2021numpyBenchmarks,
  author       = {Guelton, Serge},
  title        = {numpy-benchmarks},
  year         = {2021},
  howpublished = {\url{https://github.com/serge-sans-paille/numpy-benchmarks}}
}

@misc{guelton2025pythranTests,
  author       = {Guelton, Serge},
  title        = {Pythran tests},
  year         = {2025},
  howpublished = {\url{https://github.com/serge-sans-paille/pythran/tree/master/pythran/tests}}
}

@article{yeye-Auto-pipeline,
author = {Yang, Junwen and He, Yeye and Chaudhuri, Surajit},
title = {Auto-pipeline: synthesizing complex data pipelines by-target using reinforcement learning and search},
journal={Proceedings of the VLDB Endowment},
year = {2021},
volume = {14},
number = {11},
pages = {2563–2575}
}

@misc{Jupyter,
  title        = {Project Jupyter: Jupyter Notebooks},
  howpublished = {\url{https://jupyter.org/}},
  note         = {Accessed: 2025-10-19},
  year         = {2025}
}

@misc{NumPy,
  title        = {NumPy: The fundamental package for scientific computing with Python},
  howpublished = {\url{https://numpy.org/}},
  note         = {Accessed: 2025-10-19},
  year         = {2025}
}

@inproceedings{Wang-icse,
author = {Wang, Jiawei and Li, Li and Zeller, Andreas},
title = {Better code, better sharing: on the need of analyzing jupyter notebooks},
year = {2020},
pages = {53–56},
numpages = {4},
 booktitle={42nd International Conference on Software Engineering (ICSE-NIER)}
}

@misc{cuDF,
  title        = {cuDF: GPU DataFrame Library for Python (RAPIDS)},
  howpublished = {\url{https://rapids.ai/cudf-pandas/}},
  note         = {Accessed: 2025-10-20},
  year         = {2025}
}

@article{mozzillo2023evaluation,
  title={Evaluation of Dataframe Libraries for Data Preparation on a Single Machine},
  author={Mozzillo, Angelo and Zecchini, Luca and Gagliardelli, Luca and Aslam, Adeel and Bergamaschi, Sonia and Simonini, Giovanni},
   journal={International Conference on Extending Database Technology (EDBT)},
  pages ={337-349},
  year={2025}
}

@article{elyashiv2024pd,
  title={PD-Explain: A Unified Python-native Framework for Query Explanations Over DataFrames},
  author={Elyashiv, Itay and Gilad, Amir and Isakov, Edna and Tikochinsky, Tal and Somech, Amit},
  journal={Proceedings of the VLDB Endowment},
  volume={17},
  number={12},
  pages={4473--4476},
  year={2024}
}

@inproceedings{lu2024Xorbits,
  title = {Xorbits: Automating Operator Tiling for Distributed Data Science},
  shorttitle = {Xorbits},
  booktitle = {2024 {{IEEE}} 40th {{International Conference}} on {{Data Engineering}} ({{ICDE}})},
  author = {Lu, Weizheng and He, Kaisheng and Qin, Xuye and Li, Chengjie and Wang, Zhong and Yuan, Tao and Liao, Xia and Zhang, Feng and Chen, Yueguo and Du, Xiaoyong},
  year = {2024},
  pages = {5211--5223}
}

@inproceedings{tu2019parrot,
  title={Parrot: a python-based interactive platform for information retrieval research},
  author={Tu, Xinhui and Huang, Jimmy and Luo, Jing and Zhu, Runjie and He, Tingting},
  booktitle={Proceedings of the 42nd International Conference on Research and Development in Information Retrieval (SIGIR)},
  pages={1289--1292},
  year={2019}
}

@inproceedings{akkalyoncu2020lightweight,
  title={A lightweight environment for learning experimental IR research practices},
  author={Akkalyoncu Yilmaz, Zeynep and Clarke, Charles LA and Lin, Jimmy},
  booktitle={Proceedings of the 43rd International Conference on Research and Development in Information Retrieval (SIGIR)},
  pages={2113--2116},
  year={2020}
}

@inproceedings{macavaney2022python,
  title={A Python Interface to PISA!},
  author={MacAvaney, Sean and Macdonald, Craig},
  booktitle={Proceedings of the 45th International Conference on Research and Development in Information Retrieval (SIGIR)},
  pages={3339--3344},
  year={2022}
}

@inproceedings{macdonald2018combining,
  title={Combining terrier with Apache Spark to create agile experimental information retrieval pipelines},
  author={Macdonald, Craig},
  booktitle={Proceedings of the 41st International Conference on Research and Development in Information Retrieval (SIGIR)},
  pages={1309--1312},
  year={2018}
}

\end{document}